\documentclass{moriond}
\usepackage{amsmath}

\def\be{\begin{equation}}
\def\ee{\end{equation}}
\def\bea{\begin{eqnarray}}
\def\eea{\end{eqnarray}}

\newcommand{\Photo}{}

\begin{document}
\vspace*{4cm}
\title{Strong field tests of gravity with electromagnetic and gravitational waves}

\author{Sourabh Nampalliwar}

\address{Theoretical Astrophysics,
Eberhard Karls Universität T\"{u}bingen, T\"{u}bingen, Germany}

\maketitle\abstracts{
For nearly a century, Einstein’s theory of gravity has been the standard theory for describing gravitational phenomena in our universe. Along with its successes, limitations of the theory from theoretical (e.g., singularities) and observational (e.g., dark matter/energy) perspectives have appeared. This has led to proposals that modify or supersede Einstein’s theory, and testing these theories against data, especially in the strong-field regime, has emerged as a new paradigm in physics in recent years. Along with the completely new avenue of gravitational waves, new and improved techniques based on electromagnetic waves are being used to test general relativity (GR) ever more stringently. As the realm beyond GR is unknown, a popular approach is to look for theory-agnostic deviations from GR/predictions of GR. Here I describe how I have used gravitational waves, X-rays, and black hole shadows to constraints on some of these theory-agnostic deviations.}

\section{Introduction}

Einstein's theory of gravity, known as the theory of general relativity (GR hereafter), is the leading framework for describing gravitational phenomena at present. The theory, since its proposition in 1915, has gone through a plethora of tests and emerged as the theory most suitable for our universe. While successes abound, some features and implications of GR lead to questions on its suitability as the ultimate theory of gravity.
These questions arise from the theory (e.g., the presence of singularities, incompatibility with quantum mechanics and the hierarchy problem) as well the observation side (e.g., dark matter and dark energy). Even in the absence of the above two, tests of GR in novel scenarios are necessary to verify the applicability of GR in scenarios where it has not been verified. 

Until recently, a majority of tests of GR had been performed in the so called weak field regime~\cite{Will2014}. Only in the last few years, tests in the strong field regime have become possible, with black holes as ideal candidates for performing such tests. Three features in particular make BHs the favorites for such studies: Being the most compact objects in the universe, strong-field differences between GR and alternative theories manifest most prominently around BHs. 
Moreover, within GR, most astrophysical BHs are surprisingly simple objects -- only a few parameters describe it completely -- thus enabling smoking-gun type tests of GR. Finally, being ubiquitous in our universe provides a variety of systems for such tests.

Since the realm beyond GR is unknown, there are several possible alternatives/modifications to GR in the market, with no clear forerunner. In such a scenario, a popular approach is to look for theory-agnostic deviations away from GR, by using parametrically deformed metrics to represent BHs. Given the fact that most observations until now, both in the weak-field and the strong-field regime, have been consistent with GR, this theory-agnostic approach has the potential to guide us in the direction where the breakdown of GR, if detectable, will happen. 

\section{Methodology}
While there are several ways to observe astrophysical systems harboring BHs, some techniques in particular have emerged to the forefront when it comes to theory-agnostic strong-field tests of gravity.

\textit{Gravitational waves}: The typical astrophysical system detected with current GW detectors is BHs in a binary, releasing GWs as they spiral inwards before merging into each other. The early inspiral regime of this coalescence can be analyzed with a post-Newtonian framework, in which the two-body problem can be mapped to an effective one-body (EOB) problem. The dynamics of the EOB system are governed by an effective Hamiltonian. Assuming radiation-reaction force to follow the GR prescription, the rate of change of orbital frequency can be calculated and related to the GW emission under the stationary phase approximation. Following this strategy~\cite{Cardenas-Avendano:2019zxd}, I compute the constraints on a theory-agnostic parameter quantifying deviation from the Kerr solution of GR. 

\textit{X-ray spectroscopy}: While X-ray spectroscopy (XRS hereafter) has been used to study BHs for more than thirty years~\cite{Fabian:1989ej}, since 2017 the technique has evolved to perform tests of theories of gravity~\cite{Cao2017}. A typical astrophysical system detected with this technique includes a BH endowed with an accretion disk and a corona. Matter in the disk gets hot as it accretes, and radiates. Some of this radiation interacts with the corona and gets up-scattered to higher energies. A fraction of these high-energy photons irradiate the disk and get reflected. All this radiation, and the reflected one in particular~\cite{Reynolds:2013qqa}, carries information about the nature of the BH~\cite{Bambi2015}. 
The most advanced model currently available for analyzing the reflection component is \textsc{relxill\_nk}~\cite{relxillnk,Abdikamalov:2019yrr}, developed by my colleagues and me over the last few years. It uses the transfer function formalism~\cite{Cunningham1975} to efficiently compute the spectra for any set of parameter values. The framework has been used to study to a large number of astrophysical systems and provided some of the strongest constraints on deviations from GR~\cite{Abdikamalov:2019zfz}. 

\textit{BH imaging}: In 2019, the Event Horizon Telescope (EHT hereafter) collaboration released the first ``image'' of the supermassive BH at the center of the M$87$ galaxy~\cite{Akiyama:2019cqa}. From the image, we can infer a characteristic shadow in the middle -- region from where very little radiation arrives -- and a bright blurred ring-like feature surrounding the shadow. 
The shadow, when detectable, is an extremely clean observable for performing tests of gravity since it is independent of the astrophysical properties of the BH neighborhood. The EHT observation of the M87* image shadow provided constraints on its size and shape~\cite{Akiyama:2019eap}. I use a ray-tracing code to compute the effect of theory-agnostic deviation parameters on these observables, and use the EHT observation to calculate constraints on the theory-agnostic deviation parameters.
 
\section{Results}
While there exist many parametrically deformed metrics in literature, one of the most generic and interesting examples is the one proposed by Konoplya, Rezzolla \& Zhidenko~\cite{Konoplya2016} (KRZ hereafter). It  can be written in Boyer-Lindquist-like coordinates as 
\bea\label{eq:metric}
ds^2 &=& - \frac{N^2 - W^2 \sin^2\theta}{K^2} \, dt^2 - 2 W r \sin^2\theta \, dt \, d\phi
+ K^2 r^2 \sin^2\theta \, d\phi^2 
+ \frac{\Sigma \, B^2}{N^2} \, dr^2 + \Sigma \, r^2 \, d\theta^2.
\eea
Here, $N^2$, $W$, $K^2$ and $B^2$ encode deviations away from the Kerr metric and are functions of $r$ and $\theta$, and $\Sigma = 1 + a_*^2\cos^2{\theta}/r^2$ where $a_*$ is the dimensionless spin parameter. We focus on one of the deviation functions, $N^2$, which can be written as~\cite{Nampalliwar2021}
\bea
	N^2(r,\theta) = 1-\frac{2}{r} + \frac{a_*^2}{r^2} + \left(1-\frac{2}{r}\right)\frac{r_0^3}{r^3}\,\widetilde{A}_0(r), \quad \mathrm{ where } \quad \widetilde{A}_0 (x) = \cfrac{a_{01}}{1+\cfrac{a_{02}x}{1+\cfrac{a_{03}x}{1+\cdots}}},
\eea 
$x = 1-r_0/r$ and $r_0$ is the radius of the event horizon. In this article, we focus on the leading order deviation parameter, $a_{01}$.

\begin{figure}[!htp]
\begin{minipage}{0.5\linewidth}
\centerline{\includegraphics[width=0.99\linewidth]{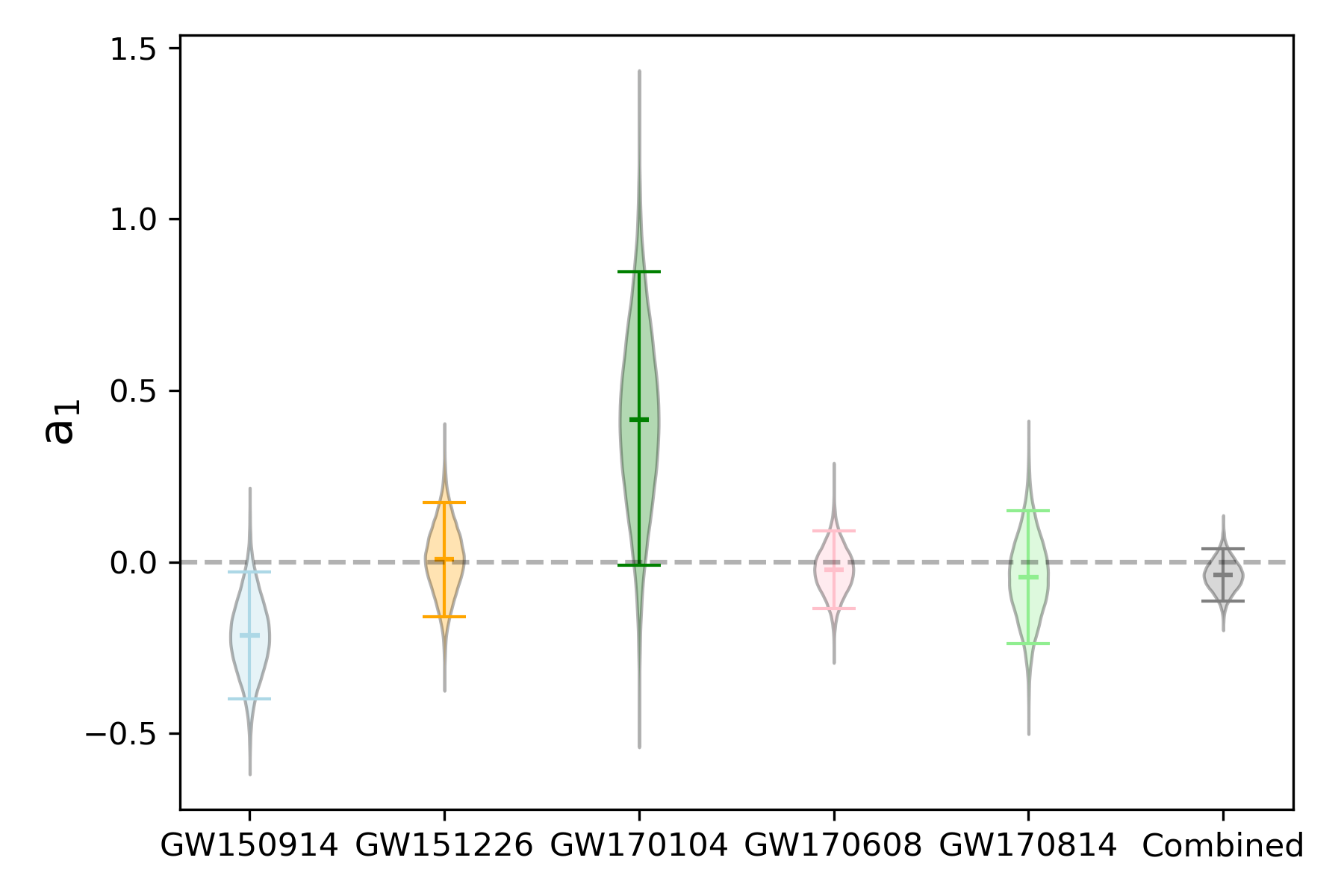}}
\end{minipage}
\hfill
\begin{minipage}{0.5\linewidth}
\centerline{\includegraphics[width=0.99\linewidth]{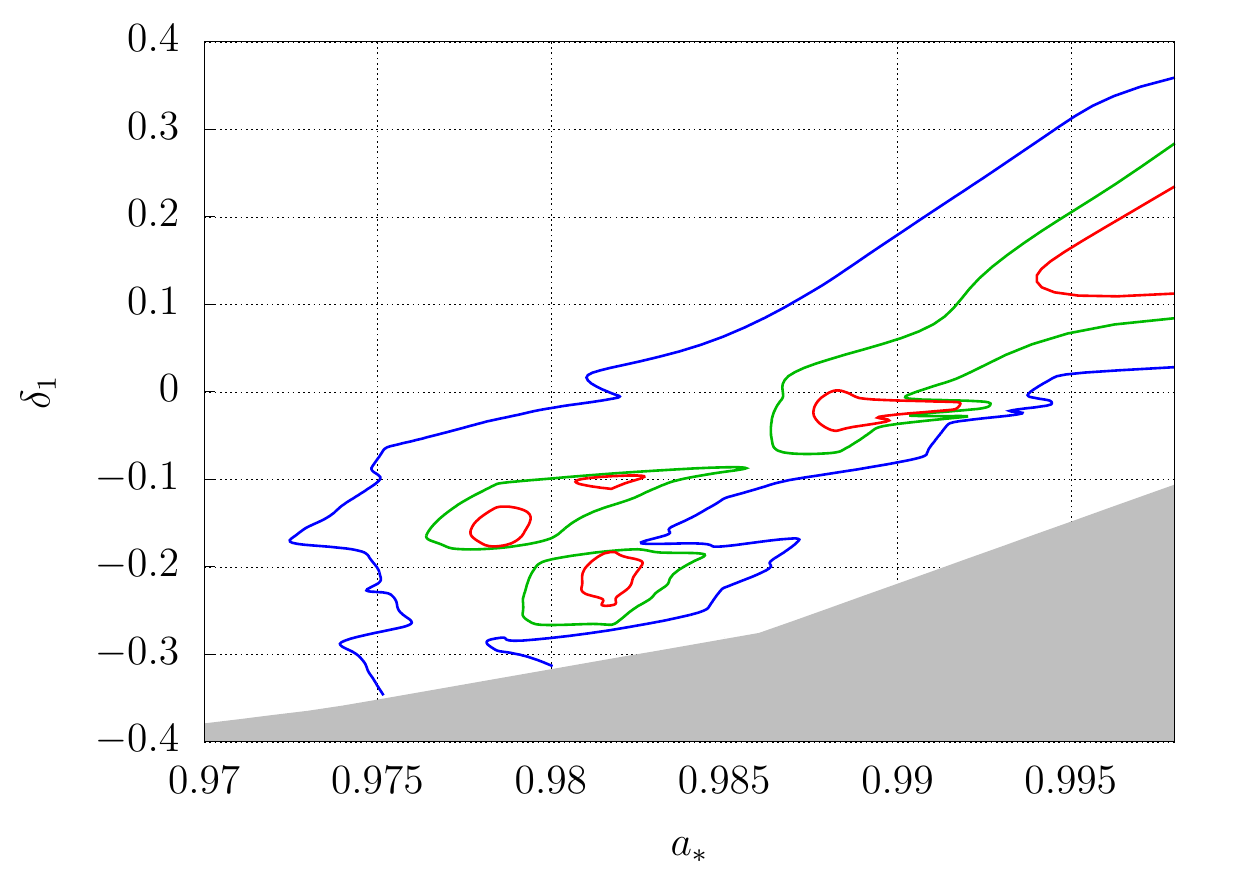}}
\end{minipage}
\caption[]{Constraints on a specific theory-agnostic deviation parameter $a_{01}$ from GWs (left panel, figure from Cardenas-Avendano, Nampalliwar \& Yunes~\cite{Cardenas-Avendano:2019zxd}) and XRS (right panel, figure from Nampalliwar, et al.~\cite{Nampalliwar:2019iti}). While the $y$-axis labels differ in each plot, they represent the same parameter which we denote by $a_{01}$. See the text for more details.
} 
\label{fig:radish1}
\end{figure}

\begin{figure}[!htp]
\centerline{\includegraphics[width=0.5\linewidth]{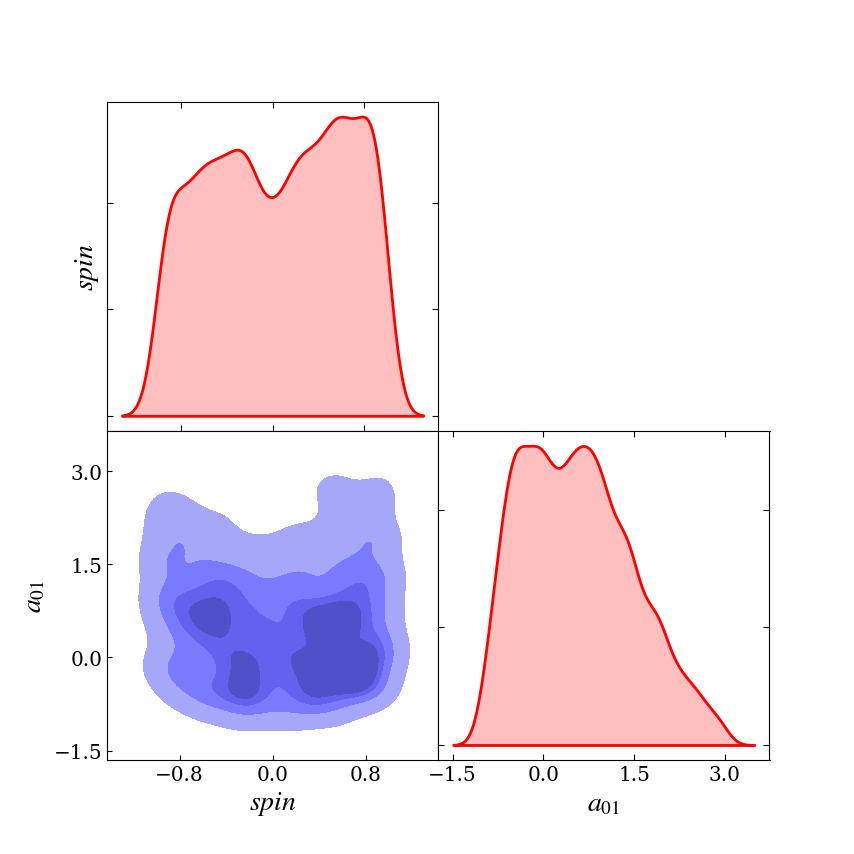}}
\caption[]{Constraints on a specific theory-agnostic deviation parameter $a_{01}$ from BH shadow. See the text for more details.} 
\label{fig:radish2}
\end{figure}

\begin{table}[!htp]
\caption[]{Constraints on a specific theory-agnostic deviation parameter $a_{01}$ from three different techniques.}
\label{tab:exp}
\vspace{0.4cm}
\begin{center}
\begin{tabular}{c|c|c|c}
Technique & Source & Lower bound & Upper bound\\
\hline
Gravitational waves & GW151226 & -0.16 & 0.17\\ \hline
X-ray reflection & Ark 564 & -0.27 & 0.28\\ \hline
BH shadow & M87* & -0.57 & 1.87
\end{tabular}
\end{center}
\end{table}

I used GWs from the strongest detections of the O1 observation run of the LIGO VIRGO GW detectors to measure $a_{01}$~\cite{Cardenas-Avendano:2019zxd}, and the results are shown in Fig.~\ref{fig:radish1} (left panel). The constraints are shown for individual sources as well as combined (the latter under the assumption that $a_{01}$ has the same value everywhere in the universe). I used the X-ray data from an active galactic nuclei, Ark 564, harboring a supermassive BH to measure $a_{01}$ with XRS~\cite{Nampalliwar:2019iti}. The results, on a spin-$a_{01}$ plot to illustrate the degeneracy between the two parameters, are shown in Fig.~\ref{fig:radish1} (right panel). The EHT observation of M87* estimated a shadow size uncertainty of 17\% and shape uncertainty $<0.05$ around its Schwarzschild value. Using this, I estimate the constraints on $a_{01}$ and the results, both 1-D and 2-D posterior distributions of $a_{01}$ and BH spin, are shown in Fig.~\ref{fig:radish2}. The constraints, at 90\% confidence level, obtained with each technique are tabulated in Table~\ref{tab:exp}. GWs provide the best constraints and XRS is comparable, while constraints from shadows are a little weaker. It is important to note here each analysis makes important assumptions which affect the constraints obtained. Instead of quantitative comparisons, I would like to highlight the qualitative advantage that a study like this can provide. While each of the technique operates on its own at present, there is significant benefit in developing synergies across the board. This can help in breaking degeneracies, check for robustness, etc.

These are early days in the field of strong-field tests of gravity. 
As models become more sophisticated, statistical and systematic uncertainties are brought under control, and better instruments become available, the field promises to provide answers to some of the biggest questions in physics.  


\section*{Acknowledgments}
I thank the Alexander von Humboldt foundation for their support.

\end{document}